\documentclass[letter,longauth]{aa} % for the letters 

\usepackage{soul} % rui

\usepackage{graphicx}
\usepackage{subfig}

\usepackage{txfonts}
\newcommand{\hii}{H~{\sc ii}}
\newcommand{\ha}{\ifmmode {\rm H}\alpha \else H$\alpha$\fi}
\newcommand{\hb}{\ifmmode {\rm H}\beta \else H$\beta$\fi}
\newcommand{\lya}{\ifmmode {\rm Ly}\alpha \else Ly$\alpha$\fi}

\newcommand{\heii}{He~{\sc ii}}
\newcommand{\Heiiuv}{He~{\sc ii} $\lambda$1640}
\newcommand{\Heiiopt}{He~{\sc ii} $\lambda$4686}

\def\msun{\ifmmode M_{\odot} \else M$_{\odot}$\fi}
\def\msunyr{\ifmmode M_{\odot} {\rm yr}^{-1} \else M$_{\odot}$ yr$^{-1}$\fi}
\def\zsun{\ifmmode Z_{\odot} \else Z$_{\odot}$\fi}
\def\lsun{\ifmmode L_{\odot} \else L$_{\odot}$\fi}

\newcommand{\oh}{\ifmmode 12 + \log({\rm O/H}) \else$12 + \log({\rm
O/H})$\fi}
\newcommand{\Civuv}{C~{\sc iv} $\lambda$1550}
\newcommand{\Civ}{C~{\sc iv}}
\newcommand{\Ciii}{C~{\sc iii}]}

\def\fesc{\ifmmode f_{\rm esc} \else $f_{\rm esc}$\fi}
\def\fesche{\ifmmode f_{\rm esc}^{\rm He+} \else $f_{\rm esc}^{\rm He+}$\fi}
\newcommand{\chion}{\ifmmode \xi_{\rm ion} \else $\xi_{\rm ion}$\fi}
\newcommand{\qrat}{\ifmmode Q_{\rm He^+}/Q_{\rm H} \else $Q_{\rm He^+}/Q_{\rm H}$\fi}

\begin{document} 

   \title{No correlation of the Lyman continuum escape fraction with spectral hardness}

\author{R.~Marques-Chaves\inst{\ref{inst1}} %\thanks{TBD}
    \and
    D.~Schaerer\inst{\ref{inst1},\ref{inst2}}
    \and
    R.~O.~Amor\'in\inst{\ref{inst3},\ref{inst4}}
    \and
    H.~Atek\inst{\ref{inst5}}
    \and
    S.~Borthakur\inst{\ref{inst6}}
    \and  
    J.~Chisholm\inst{\ref{inst7}}
    \and 
    V.~Fern\'andez\inst{\ref{inst4}}
    \and
    S.~R.~Flury\inst{\ref{inst8}}
    \and 
    M.~Giavalisco\inst{\ref{inst8}}
    \and
    A.~Grazian\inst{\ref{inst9}}
    \and
    M.~J.~Hayes\inst{\ref{inst10}}
    \and
    T.~M.~Heckman\inst{\ref{inst11}}
    \and
    A.~Henry\inst{\ref{inst12}}
    \and
    Y.~I.~Izotov\inst{\ref{inst13}}
    \and
    A.~E.~Jaskot\inst{\ref{inst14}}
    \and
    Z.~Ji\inst{\ref{inst8}}
    \and 
    S.~R.~McCandliss\inst{\ref{inst11}}
    \and
    M.~S.~Oey\inst{\ref{inst15}}
    \and
    G.~\"Ostlin\inst{\ref{inst10}}
    \and
    S.~Ravindranath\inst{\ref{inst12}}
    \and 
    M.~J.~Rutkowski\inst{\ref{inst16}}
    \and 
    A.~Saldana-Lopez\inst{\ref{inst1}}
    \and
    H.~Teplitz\inst{\ref{inst17}}
    \and
    T.~X. Thuan\inst{\ref{inst18}}
    \and
    A.~Verhamme\inst{\ref{inst1}}
    \and
    B.~Wang\inst{\ref{inst11}}
    \and
    G.~Worseck\inst{\ref{inst19}}
    \and
    X.~Xu\inst{\ref{inst11}}
}

\institute{
    Department of Astronomy, University of Geneva, 51 Chemin Pegasi, 1290 Versoix, Switzerland \label{inst1}
    \and
    CNRS, IRAP, 14 Avenue E. Belin, 31400 Toulouse, France \label{inst2}
    \and
    Instituto de Investigaci\'on Multidisciplinar en Ciencia y Tecnolog\'ia, Universidad de La Serena, Ra\'ul Bitr\'an 1305, La Serena, Chile \label{inst3}
    \and
    Departamento de F\'isica y Astronom\'ia, Universidad de La Serena, Avda. Juan Cisternas 1200, La Serena, Chile \label{inst4}
    \and
    Institut d’astrophysique de Paris, CNRS UMR7095, Sorbonne Universit\'e, 98bis Boulevard Arago, F-75014 Paris, France\label{inst5}
    \and 
    School of Earth \& Space Exploration, Arizona State University, Tempe, AZ 85287, USA \label{inst6}
    \and
    Department of Astronomy, The University of Texas at Austin, 2515 Speedway, Stop C1400, Austin, TX 78712-1205, USA \label{inst7}
    \and 
    Department of Astronomy, University of Massachusetts, Amherst, MA 01003, USA \label{inst8}
    \and
    INAF-Osservatorio Astronomico di Padova, Vicolo dell' Osservatorio, 5, 35122 Padova, Italy \label{inst9}
    \and
    Department of Astronomy, Oskar Klein Centre; Stockholm University; SE-106 91 Stockholm, Sweden \label{inst10}
    \and    
    Department of Physics and Astronomy, Johns Hopkins University, 3400 North Charles Street, Baltimore, MD 21218, USA \label{inst11}
    \and   
    Space Telescope Science Institute; 3700 San Martin Drive, Baltimore, MD, 21218, USA \label{inst12}
    \and
    Bogolyubov Institute for Theoretical Physics, National Academy of Sciences of Ukraine, 14b Metrolohichna str., Kyiv, 03143, Ukraine \label{inst13}
    \and
    Astronomy Department, Williams College, Williamstown, MA 01267, USA \label{inst14}
    \and     
    University of Michigan, Department of Astronomy, 323 West Hall, 1085 S. University Ave, Ann Arbor, MI 48109, USA \label{inst15}
    \and
    Department of Physics and Astronomy, Minnesota State University, Mankato, MN, 56001, USA\label{inst16}
    \and
    Infrared Processing and Analysis Center, California Institute of Technology, Pasadena, CA 91125, USA \label{inst17}
    \and
    Astronomy Department, University of Virginia, P.O. Box 400325, Charlottesville, VA 22904-4325, USA \label{inst18}
    \and
    Institut für Physik und Astronomie, Universität Potsdam, Karl-Liebknecht-Str. 24/25, D-14476 Potsdam, Germany\label{inst19}
}

   \date{Received --; accepted --}
 
   \abstract

\abstract{The 
%\LEt{questions should be avoided in academic writing, therefore I suggest a rephrase of your title. It does not imply that there really is a correlation, even if it isn't phrased as a question}
properties that govern the production and escape of hydrogen-ionizing photons (Lyman continuum, LyC; with energies >13.6 eV) in star-forming galaxies are still poorly understood, but they are key to identifying and characterizing the sources that reionized the Universe. 
Here we empirically explore the relation between the hardness of ionizing radiation and the LyC leakage in a large sample of low-$z$ star-forming galaxies from the recent {\em Hubble Space Telescope}  Low-$z$ Lyman Continuum Survey. 
Using Sloan Digital Sky Survey stacks and deep XShooter observations, we investigate the hardness of the ionizing spectra (\qrat) between 54.4~eV (He$^{+}$) and 13.6 eV (H) from the optical recombination lines \heii~4686\AA\ and \hb~4861\AA\ for galaxies with LyC escape fractions spanning a wide range, $f_{\rm esc} \rm (LyC) \simeq 0 - 90\%$. We find that the observed intensity of \heii/\hb\ is primarily driven by variations in the metallicity, 
but is not correlated with LyC leakage. Both very strong (<$f_{\rm esc} \rm (LyC)$>~$\simeq 0.5$) and nonleakers (<$f_{\rm esc} \rm (LyC)$>~$\simeq 0$) present similar observed intensities of \heii\ and \hb\ at comparable metallicity, between $\simeq 0.01$ and $\simeq 0.02$ for \oh~$>8.0$ and $<8.0$, respectively. 
Our results demonstrate that \qrat\ does not correlate with $f_{\rm esc} \rm (LyC)$, which implies that strong LyC emitters do not show harder ionizing spectra than nonleakers at similar metallicity. 
}

    \maketitle
%
%-------------------------------------------------------------------

\section{Introduction}

Numerous studies of star-forming galaxies at $z \sim 1-3$ have shown differences with low-$z$ galaxies in the Sloan Digital Sky Survey (SDSS), based on shifts in the classical, optical BPT emission-line diagrams  toward {\em \textup{higher excitation}} in the nebular gas phase \citep[see the review of ][and references therein]{Kewley2019Understanding-G}.
Such shifts can be due to different physical effects causing harder ionizing spectra of their stellar populations (e.g.,\ lower metallicities, younger ages), interstellar medium (ISM) properties (higher ionization parameter, different gas pressure and/or density), $\alpha$-element enhancements, selection effects, or others 
\citep[cf.][]{Steidel2016,Kewley2019Understanding-G,Izotov2021Low-redshift-co}.

Possibly larger differences have been observed in rest-UV spectra of distant galaxies, where strong nebular emission from 
\Civuv\ and \Heiiuv\ has been found \citep[e.g.,][]{Stark2015Spectroscopic-d,Mainali2017Evidence-for-a-,Vanzella2021The-MUSE-Deep-L}.
These lines have also been detected in confirmed or suspected Lyman-continuum  (LyC) emitters at $z \sim 2-4$  \citep[see][]{vanzella2018,Vanzella2020Ionizing-the-in,Naidu2021}. 
These observations suggest hard ionizing spectra in particular because \Civ\ and \heii\ probe energies above 47.8 and 54.5~eV, respectively, and are thus sensitive to higher-energy radiation than the classical strong optical emission lines of [O~{\sc iii}], [O~{\sc ii}], [N~{\sc ii}], and [S~{\sc ii}]
\citep[e.g.,][]{Kewley2019Understanding-G,Feltre2016}. 

Nebular \heii\ emission provides the best measure of the hardness of the ionizing radiation field because its recombination lines\footnote{The strongest and most commonly detected lines are \Heiiuv\ and \Heiiopt.} are basically direct photons counters for energies $>54$ eV, whereas the forbidden metal lines depend on many parameters (ionization parameter and others).
\cite{Naidu2021} stacked rest-UV spectra of Ly$\alpha$ emitters (LAEs) at $z \sim 2,$ for which they estimated LyC escape fractions using indirect methods, finding narrow  \Heiiuv, \Civuv, and other lines in sources with high LyC escape, whereas low escape sources only show \Ciii\ and O~{\sc iii}] emission.
Based on this, they suggested a possible relation between the hardness of the ionizing spectra and LyC escape.
\heii\ is also seen in the stacks of the LyC candidates of \cite{Marchi2018Lyalpha-Lyman-c} selected as Lyman break galaxies (LBGs), although the poor resolution makes it difficult to exclude a significant contribution from stellar emission. Moreover, nebular \heii\ emission has not been reported in the LBG stacked LyC emitter spectra of \cite{Steidel2018}. 
These findings call for a clarification of the possible link between LyC escape, nebular \heii\ emission, and the hardness of the radiation between energies 13.6 and 54 eV.

At low redshift, \cite{Jaskot2013The-Origin-and-} have examined how hard ionizing radiation can influence emission line diagnostics of the optical depth of LyC radiation in green pea galaxies, a class of strong emission line galaxies now known to contain LyC emitters.
\cite{Schaerer2022} have recently discovered intense \Civ\ and \Heiiuv\ emission lines in three low-$z$ strong LyC emitters, ($\fesc > 0.1$), with UV properties similar to the high-$z$ galaxies mentioned above. They proposed that strong \Civuv\ emission indicates high LyC escape fractions. 
Furthermore, they also estimated that strong LyC leakers do not have harder ionizing spectra than nonleakers, and that the presence of strong \Civ\ and \heii\ in the spectra of LyC leakers could be primarily due to a high ionizing photon production \citep{Schaerer2022}.

On the other hand, \cite{Perez-Montero2020Photon-leaking-} suggested that the observed \Heiiopt\ emission in metal-poor low-$z$ galaxies could be explained by significant photon leakage, without the need for additional sources of hard ionizing photons, while it is generally accepted that normal stellar populations cannot produce sufficient photons above 54 eV \citep[see,
%\LEt{you use "cf." and "see". Please decide for one and change to this throughout for consistency} 
e.g.,][]{Stasinska2015Excitation-prop,schaerer2019,olivier2021, simmonds2021}. 
These results again call for a systematic approach to determine whether LyC emitting galaxies show a different He$^+$-ionizing radiation than comparable galaxies with negligible Lyman continuum escape.

In this Letter we determine empirically and in a differential manner the hardness of the ionizing spectra between 54.4 and 13.6 eV (i.e.,\ the ionization potential of He$^+$ and that of H)
as a function of the LyC escape fraction for the first time. 
To achieve this goal, we use the data from 89 low-redshift star-forming galaxies for which
Lyman continuum and  nonionizing UV spectra are available from the recent {\em Hubble Space Telescope} ({\em HST}) Low-$z$ Lyman Continuum Survey 
and other observations (see Sect.\ 2). 
Using optical spectra and stacking, we determine the hardness of ionizing radiation field from 
the optical recombination lines \Heiiopt\ and \hb\ in a straightforward manner, in contrast to methods using forbidden optical lines and to
methods relying on UV emission lines. 
Our first constraints on the hardness of LyC emitters and a control sample thus contribute to a better knowledge of the analogue sources of cosmic reionization, their ionizing radiation field and ISM.

%--------------------------------------------------------------------
\section{Observations}

\subsection{Lyman continuum and UV observations}

We used the recent LzLCS, which consists of 66 star-forming galaxies at $z \simeq 0.2-0.4$ \citep{Flury2022a,Flury2022b,Saldana2022} and 23 additional sources previously studied by \cite{izotov2016a, izotov2016b, izotov2018a, izotov2021}  and \cite{wang2019}. 
This unique dataset includes 37 LyC emitters detected with high signal-to-noise ratio (S/N>3), 
%\LEt{please change to "S/N" throughout to avoid confusion with "SNR" for "supernova remnant"}>3),
with escape fractions spanning a wide range (\fesc $\sim$ 1--90 \%), and strong upper limits for the remaining sources.
The galaxies span a wide range of physical properties with stellar masses $\log(M_{*}/M_{\odot})=7.2-10.8$, star formation rates SFR$=(3-80)$~$M_{\odot}$~yr$^{-1}$, and nebular abundances $\oh = 7.5 - 8.6$, derived using the direct method \citep{Flury2022a}.

\subsection{SDSS optical spectra}\label{sec_21}

Optical spectra are available for all sources from SDSS and BOSS surveys \citep{eisenstein2011}.
We retrieved optical spectra from the SDSS Science Archive Server. 
SDSS spectra have a spectral resolution $R\sim 2000$. 
The extinction curve of \cite{fitzpatrick1999} and the Galactic $E(B-V)$ values from the dust maps of \cite{green2018} were used to correct the reddening effect of the Galaxy. 
Flux measurements were derived using Gaussian profiles and the {\sc Python} nonlinear least-squares function {\sc curve\_fit} and corresponding uncertainties using a Monte Carlo method. The \cite{cardelli1989} reddening law ($R_{V}=3.1$) was adopted to correct for the internal extinction using the ratios of well-detected Balmer emission lines \citep[following][]{izotov1994}.

In the lines of interest in this work, H$\beta$ is detected with high significance in all galaxies, but not He~{\sc ii}, yielding $3\sigma$ limits of He~{\sc ii}/H$\beta \lesssim 0.04$ on average, which are fairly above the typical intensities He~{\sc ii}/H$\beta \lesssim 0.02$ found in other compact star-forming galaxies \citep[e.g.,][]{izotov2016c}, including LyC emitters \citep{guseva2020}. Therefore, we performed a stacking analysis of the SDSS spectra to improve the He~{\sc ii}/H$\beta$ limits.

\begin{table*}
\caption{Summary of the results of SDSS stacks}
\label{table1} 
\centering 
\resizebox{\textwidth}{!}{%
\begin{tabular}{l l l c c c c c}          
\hline\hline 
SDSS Bins  & Denomination & Definition & N & \fesc\  & \oh & He~{\sc ii}/H$\beta$ & \qrat \\ 
(1) & (2) & (3) & (4) & (5) & (6) & (7) &  (8) \\
\hline   
\fesc\ & very strong leakers  & $\fesc \geq 0.2$ & 8 & $0.497_{-0.140}^{+0.242}$ & $7.849_{-0.165}^{+0.265}$ & $0.016 \pm 0.007$ & $0.005 \pm 0.002$\\

 & strong leakers & $0.05 \leq \fesc\  < 0.2$ & 13 & $0.117_{-0.042}^{+0.053}$ & $8.107_{-0.221}^{+0.296}$ & $0.013 \pm 0.006$ & $0.007 \pm 0.003$\\
 
 & weak leakers & $\fesc\ < 0.05$ & 29 & $0.025_{-0.011}^{+0.014}$ & $8.115_{-0.174}^{+0.221}$ & $0.012 \pm 0.004$ & $0.007 \pm 0.002$\\

 & non-leakers & $\rm S/N (LyC) < 2$ & 39 & $<0.011$ & $8.175_{-0.186}^{+0.245}$ & $0.015 \pm 0.005$ & $0.009 \pm 0.003 $\\
\hline   

\fesc\ and metallicity & & $\fesc \geq 0.05$ and \oh~< 8.0 & 11 &  $0.346_{-0.228}^{+0.
322}$ & $7.811_{-0.109}^{+0.161}$ & $0.020 \pm 0.005$ & $0.008 \pm 0.002$\\

 & & $\fesc \geq 0.05$ and \oh~$= 8.0-8.3$ & 6 & $0.203_{-0.113}^{+0.159}$ & $8.095_{-0.086}^{+0.110}$ & $0.013 \pm 0.006$ & $0.006 \pm 0.003$ \\

 & & $\fesc \geq 0.05$ and \oh~> 8.3 & 4 & $0.120_{-0.056}^{+0.066}$ & $8.423_{-0.051}^{+0.102}$ & $<0.011$ & $< 0.006$\\

%\hline

 & & $\fesc < 0.05$ and \oh~< 8.0 & 9 & $0.030_{-0.010}^{+0.013}$ & $7.895_{-0.042}^{+0.060}$ & $<0.016$  & $<0.009$\\

 & & $\fesc < 0.05$ and \oh~$= 8.0-8.3$ & 15 & $0.022_{-0.011}^{+0.015}$ & $8.153_{-0.084}^{+0.100}$ & $0.015 \pm 0.004$ & $0.008\pm0.002$ \\

 & & $\fesc < 0.05$ and \oh~> 8.3 & 5 & $0.023_{-0.009}^{+0.015}$ & $8.395_{-0.029}^{+0.122}$ & $<0.023$ & $<0.013$ \\ 

%\hline

 & & $\rm S/N (LyC) < 2$ and \oh~< 8.0 & 6 & $<0.027$ & $7.786_{-0.120}^{+0.173}$ & $0.030 \pm 0.009$ & $0.017\pm0.005$ \\

 & & $\rm S/N (LyC) < 2$ and \oh~$= 8.0-8.3$ & 22 & $<0.008$ & $8.175_{-0.075}^{+0.091}$ & $0.011 \pm 0.004$ & $0.006\pm0.002$  \\
 & & $\rm S/N (LyC) < 2$ and \oh~> 8.3 & 11 & $<0.009$ & $8.388_{-0.057}^{+0.086}$ & $< 0.013$ & $<0.007$  \\

\hline 
\end{tabular} }
\textbf{Notes. } Columns (1), (2), and (3): Groups, denomination, and definition of each bin. Column (4): Number of sources in each bin. Columns (5) and (6): mean and 68\% confidence intervals (bootstrap resampling) of \fesc\ and \oh\ for each bin. Column (7): Observed \heii/\hb\ intensities. Column (8): Hardness of the ionizing spectra between 54.4~eV (He$^{+}$) and 13.6 eV (H) following Eq.~\ref{eq_1} and assuming \fesche~$\approx 0$. Upper limits refer to a $2\sigma$ limit.
\end{table*}

We built two different groups of stacked spectra, one composed of bins in absolute LyC escape fraction, $f_{\rm esc} \rm (LyC)$, and another one composed of bins in $f_{\rm esc} \rm (LyC)$ and \oh. 
For the first group, we defined the four bins using \fesc\ determined from the UV analysis of \cite{Saldana2022}. For the second group, we built nine stack spectra considering three bins of $f_{\rm esc} \rm (LyC)$ and three bins of metallicity. Table \ref{table1} contains the definition of the bins and the main derived properties from the stacked spectra.

\begin{figure}
  \centering
  \includegraphics[width=0.46\textwidth]{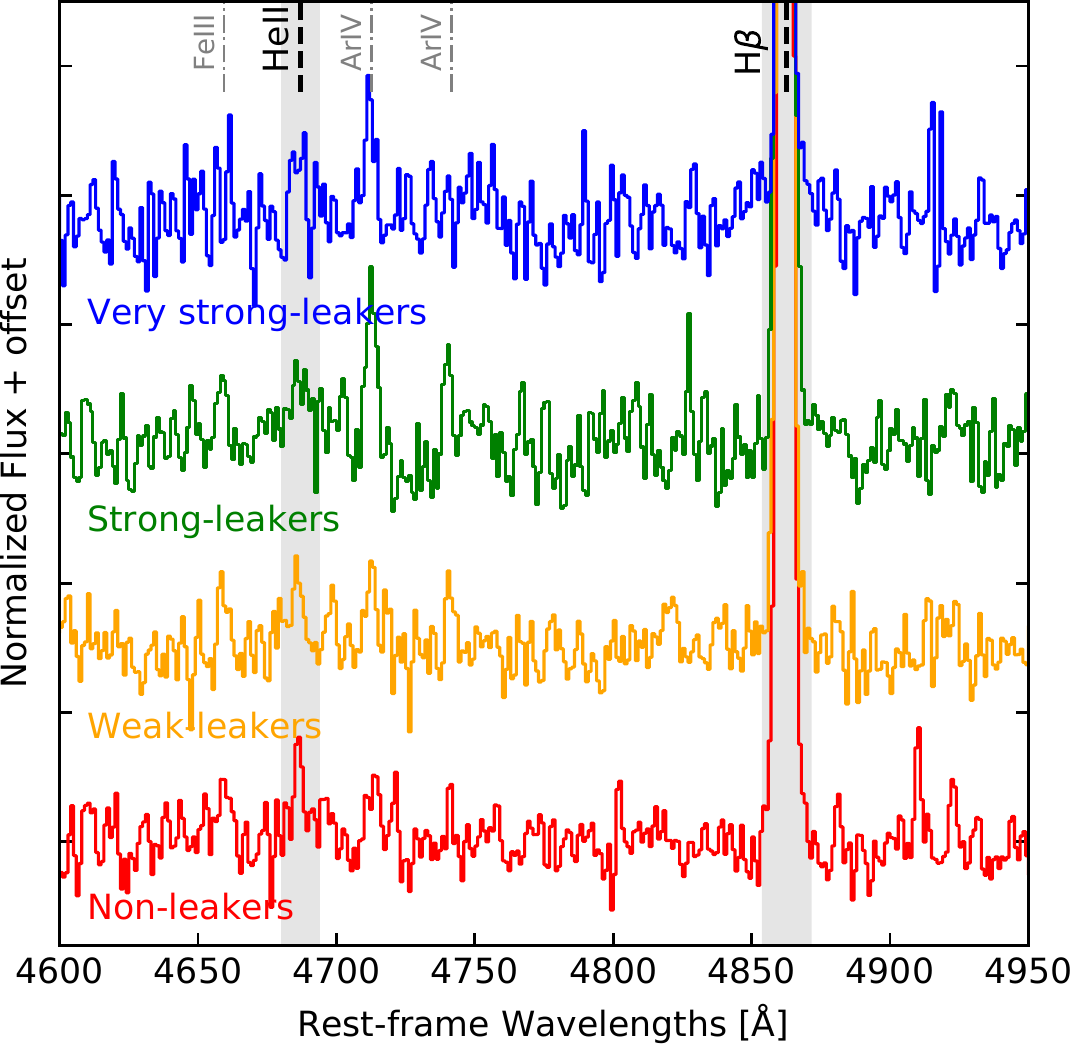}
  \caption{Normalized stacked SDSS spectra in bins of \fesc : Nonleakers (red), weak leakers (orange), strong leakers (green), and very strong leakers (blue). Regions around He~{\sc ii}~4686\AA{ }and H$\beta$~4861\AA{ }are highlighted in gray. He~{\sc ii} is detected in the four stacks with $\rm S/N \sim 2-3$. Other faint lines are also detected ([Fe~{\sc iii}]~4658\AA{ }, [Ar~{\sc iv}]+He~{\sc i}~4712\AA,{ }and [Ar~{\sc iv}~4740\AA), and their positions are marked with dashed gray lines.}
  \label{fig1}
\end{figure}

For each bin, the SDSS spectra were deredshifted using the systemic redshifts from the observed wavelengths of bright optical lines and were resampled using a linear interpolation onto a common wavelength grid. Next, we normalized the spectra at $\lambda = 4750-4850$\AA, that is, relatively free of emission or absorption features. Finally, we stacked all spectra by averaging the flux in each spectral bin. We also tested other stacking methods using the median and the weight-average using the uncertainty spectra, but no significant differences in the observed intensity of He~{\sc ii}/H$\beta$ were found between these three methods. Fig.~\ref{fig1} shows the four stacked spectra in bins of $f_{\rm esc} \rm (LyC)$. He~{\sc ii} is detected at least in three stacks, together with other faint lines, such as [Fe~{\sc iii}]~4658\AA{ }and the [Ar~{\sc iv}] doublet at 4711\AA{ }and 4740\AA. We note the lack of significant broad emission in the stacks  around 4650\AA\ and 5808\AA, which might be associated with a significant contribution of Wolf-Rayet  stars \citep[e.g.,][]{brinchmann2008b}.

\subsection{XShooter/VLT observations}

Spectroscopic observations of eight LzLCS sources were carried out during 2021 with the XShooter instrument on the Very Large Telescope (VLT), with total integration times of 50-100 min.
The XShooter spectra were reduced in a standard manner using the ESO Reflex reduction pipeline (version 2.11.5; \citealt{freudling2013}) to produce flux-calibrated spectra. The Galaxy and internal reddening were corrected using the same method as described in Section \ref{sec_21}. We also used the measurements for five other LyC emitters observed with XShooter by \cite{guseva2020}. In total, He~{\sc ii} is detected in nine sources at >$2\sigma$, with intrinsic He~{\sc ii}/H$\beta \approx 0.007-0.020$. For the remaining four sources, $2\sigma$ upper limits between $\simeq 0.004$ and $\simeq 0.010$ are inferred.

\section{Results and discussion}

%We now examine the resulting \heii/\hb\ measurements and the inferences on the hardness of the ionizing spectra as a function of Lyman continuum escape and other parameters\LEt{a single sentence does not constitute a paragraph. Please either add to this or remove}.

\subsection{HeII/H$\beta$ versus escape of LyC photons}

\begin{figure*}
  \centering
  \includegraphics[width=0.97\textwidth]{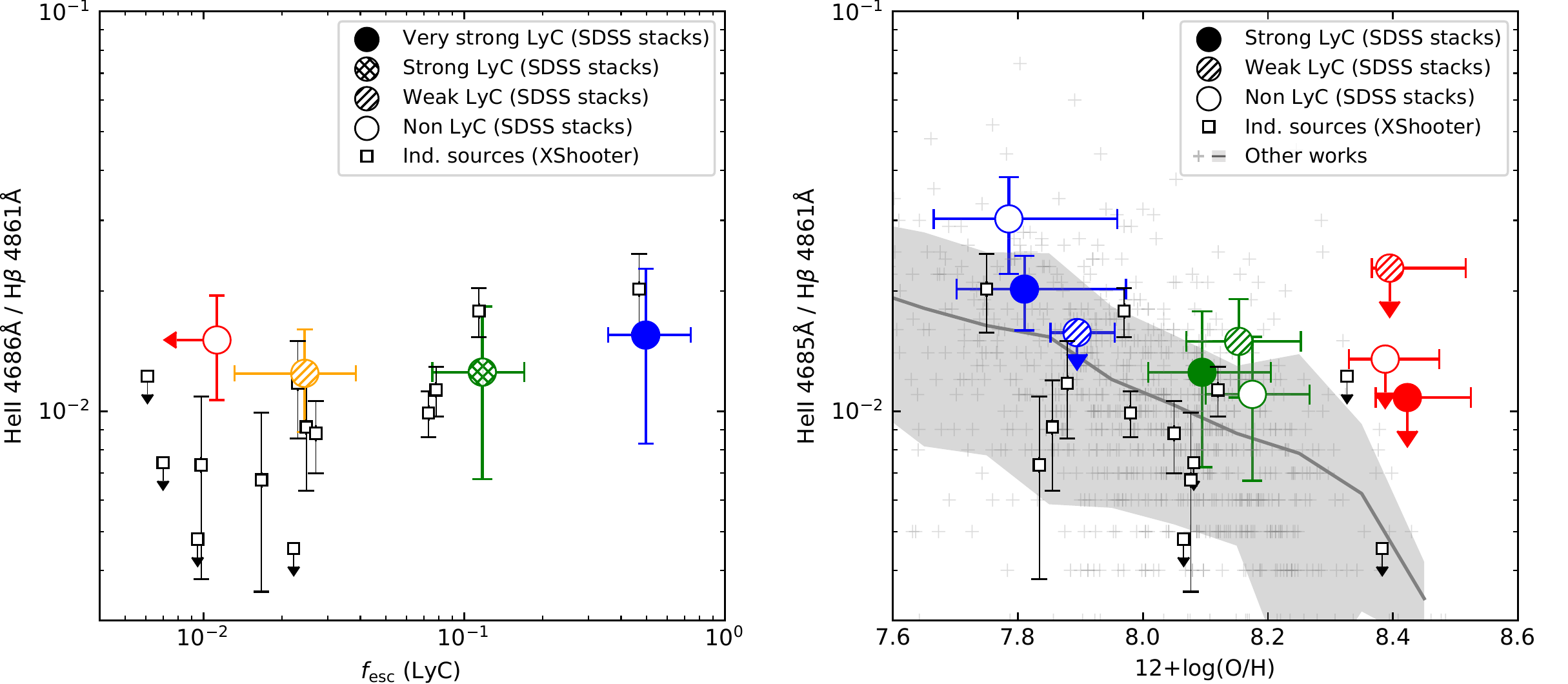}
  \caption{Intensity of He~{\sc ii}/H$\beta$ as a function of $f_{\rm esc} \rm (LyC)$ (left) and \oh\ (right) for different SDSS stacks (circles). The left panel shows the four stacks in bins of $f_{\rm esc} \rm (LyC)$ (very strong, strong, weak, and nonleakers in blue, green, yellow, and red, respectively), and the right panel shows the nine stacks in bins of $f_{\rm esc} \rm (LyC)$ and \oh (\oh<8.0, 8.0-8.3, and $>$8.3, in blue, green, and red, respectively). Details of the definition of stacks can be found in Table \ref{table1}.
  $x$-values and horizontal error bars refer to the mean and 68\% confidence intervals (bootstrap resampling) of $f_{\rm esc} \rm (LyC)$ and \oh\ measured for each bin (Table \ref{table1}). Individual sources with XShooter observations are also shown (empty squares; LzLCS sources and five LyC emitters were analyzed in \citealt{guseva2020}). Upper limits for SDSS stacks and XShooter observations refer to $2\sigma$ limits. 
  The right panel also shows observations (plusess) of a compilation of almost 900 star-forming galaxies from \cite{izotov2016c} (average and standard deviation marked with a solid line and region, respectively). %\LEt{the caption to a figure should only contain information required to understand the figure, no interpretation. To my understanding, the next sentences should be removed. Please consider this }
%  No significant evolution of He~{\sc ii}/H$\beta$ with $f_{\rm esc} \rm (LyC)$ is found. Strong LyC leakers present roughly the same \heii/\hb\ intensities as in weak or nonleakers at comparable metallicity, and these do not differ from those observed in other star-forming galaxies. He~{\sc ii}/H$\beta$ is primarily driven by variations on the metallicity, where He~{\sc ii}/H$\beta$ increases with decreasing \oh, in particular for \oh<8.0. 
}
  \label{fig2}
\end{figure*}

The main result of our study is presented in the left panel of Fig.~\ref{fig2}, which shows the relation between the observed He~{\sc ii}/H$\beta$ ratio and the observed LyC escape fraction \fesc.
Overall, we find no significant dependence of the \heii/\hb\ intensity on \fesc\ from the SDSS stacks.

To quantify this, we computed {\it \textup{Kendall's}}-$\tau$ correlation coefficient between these two variables following \cite{akritas1996}, which allows the inclusion of upper limits. Using the results from SDSS stacks, we find $\tau (\rm SDSS) = 0.00^{+0.33}_{-0.50}$ ($p$-$value = 1$, uncertainties derived using a Monte Carlo method), which is basically consistent with a null correlation within our uncertainties. On the other hand, using the results from the 13 individual sources with deep XShooter spectra, we find a positive correlation with $\tau (\rm XS) = 0.603^{+0.103}_{-0.128}$ ($p$-$value = 4.14\times 10^{-3}$). 
However, these results are statistically less meaningful than those obtained with SDSS stacks using the average properties of 89 sources that also include the XShooter sources. In addition, the underlying effect of \oh\ on \heii/\hb\ has not been considered so far.

The intensity of \heii/\hb\ of star-forming galaxies is known to vary with metallicity in that it increases with decreasing of \oh\ \citep[e.g.,][]{brinchmann2008b, shirazi2012, schaerer2019}.
As our sample spans a wide range of metallicities, $\oh = 7.5 - 8.6$ \citep{Flury2022a}, this behavior should be considered in our analysis.
Therefore, we used nine SDSS stacks grouped in three bins of \fesc\ and three bins of \oh~(see Table \ref{table1} for details).

The right panel of Fig.~\ref{fig2} shows the relation between \heii/\hb\ and \oh\ for strong LyC leakers (solid circles) and non- or weak LyC leakers (empty and dashed circles) in the metallicity bins. For comparison, we also show the observed intensities of \heii/\hb\ and metallicities of a compilation of almost 900 star-forming galaxies from SDSS DR14 \citep[pluses;][]{izotov2016c}, for which the [O~{\sc iii}]~4363\AA{ }line is detected with an accuracy better than 4$\sigma$, allowing thus direct abundance determinations using the $T_{\rm e}$ method. The overall properties of the parent sample are discussed in \cite{guseva2019}.

Two main conclusions can be drawn from the right panel of Fig.~\ref{fig2}. First, strong LyC leakers present roughly the same \heii/\hb\ intensities as non- or weak leakers at comparable metallicities. 
For instance, strong \heii/\hb\  intensities are found for both nonleakers and strong LyC leakers in the low-metallicity bin (\heii/\hb~$= 0.030 \pm 0.009$ and $0.020\pm 0.005$, respectively), while sources with $\oh = 8.0-8.3$ show \heii/\hb~$\sim 0.012$, independently of $f_{\rm esc} \rm (LyC)$.
Moreover, for a specific range of \oh, the line intensities \heii/\hb\ inferred for strong leakers do not differ from those typically observed in other star-forming galaxies \citep{izotov2016c}. 
Second, low-metallicity star-forming galaxies present stronger \heii/\hb\ than high-metallicity galaxies. This
trend has been reported in other works of increasing the \heii/\hb\ intensity with decreasing metallicity. The same trend is seen in the individual sources with XShooter observations, where we find an anticorrelation between \heii/\hb\ and metallicity of $\tau (\rm XS) = -0.487^{+0.103}_{-0.103}$ ($p$-$value = 2.04\times 10^{-2}$). For example, the two XShooter sources with the highest \heii/\hb\ ratios ($\simeq 0.02$) present \oh <8.0, while the sources that are not detected in He~{\sc ii} all have \oh >8.0.
This behavior explains the positive correlation between \heii/\hb\ and \fesc\ seen in the individual sources with deep XShooter spectra (Fig.~\ref{fig2}, left).

\subsection{No variation in hardness with LyC escape}

The hardness of the ionizing radiation field between energies above 54.4 and 13.6 eV, described by \qrat, is to first order related to the relative recombination line intensities by
\begin{equation}
    \begin{aligned}
        I(4686)/I(\hb)& = \frac{c_{4686}}{c_{\hb}} \frac{(1-\fesche) \int_{54.4}^\infty (F_\nu/h \nu) \, d\nu}{(1-\fesc) \int_{13.6}^\infty (F_\nu/h \nu) \, d\nu}
        \\ \noalign{\vskip5pt}
       & = 1.74 \frac{(1-\fesche) \, Q_{\rm He^+}}{(1-\fesc) \, Q_{\rm H}},
    \end{aligned}
    \label{eq_1}
\end{equation}
where $Q$ expresses the number of ionizing photons emitted above the corresponding ionization potential, and $c_{ji} = h \nu_{ji} \frac{\alpha^{\rm eff}_{ji}}{\alpha_B}$ relates the recombination rate to the line intensity \citep{osterbrock2006}. For $c_{ji}$ we have adopted typical values of the electron temperature ($T_e=10$ kK). 
The above expression also accounts for the escape of ionizing photons, which are \emph{\textup{a priori}} different for He$^+$ and H-ionizing photons. It is expected that $\fesche \ll \fesc$ because the doubly ionized He region is generally significantly smaller than the \hii\ region, or in other words, He$^+$-ionizing photons are absorbed much closer to the source than those of lower energy. Except possibly for extremely hard, power-law-like ionizing spectra, \fesche\ is therefore expected to be very low, or at least lower than \fesc.

We have shown that the intensity of \heii/\hb\ does not depend on \fesc and is primarily driven by variations with metallicity. 
Therefore, Eq.~\ref{eq_1} implies that the hardness \qrat\ does not correlate with the LyC escape, since, if anything, \heii/\hb\ should increase with increasing \fesc\ even for constant \qrat\ (see Eq.~\ref{eq_1}). Table~\ref{table1} provides the inferred values of \qrat\ for all the stacks using Eq.~\ref{eq_1} and assuming \fesche~$\approx 0$. No variation is found in \qrat\ between strong and nonleakers within the uncertainties. 

\subsection{Implications}

Our results rule out LyC leakage to explain the origin of nebular \heii\ emission put forward by \cite{Perez-Montero2020Photon-leaking-}. 
Their galaxies present roughly the same \heii/\hb\ intensities at comparable metallicity as those studied in our work (see their Figure~1). 
According to these authors, the observed intensities of He~{\sc ii} could be explained by density-bounded H~{\sc ii} regions with very highly ionizing photon leaking, with a mean $f_{\rm esc} \rm (LyC) \simeq 0.74$ for the entire sample. Clearly, the results shown in Fig.~\ref{fig2} contradict this scenario, where the same intensities of \heii/\hb\ are found in sources with both very high and very low $f_{\rm esc} \rm (LyC)$, that is, without any dependence on \fesc. 
Furthermore, the invoked mean $\fesc \simeq 0.74$ is significantly higher than measured in comparable low-$z$ galaxies
\citep{Flury2022a,Flury2022b}, and for the strongest known low-$z$ leaker (J1243$+$4646 with $\fesc \simeq 0.73$ reported by \cite{izotov2018b}, or $\fesc \simeq 0.89$ inferred in 
%\LEt{to properly include this reference into the main text, please add an opening parenthesis before the year. This is a LaTeX command error that I cannot fix for you with the program I work with (citet and citep)}
\citet{Saldana2022}), we measure \heii/\hb~$<0.02$ (2$\sigma$), which is not exceptionally high compared to many other star-forming galaxies with similar metallicity (\oh~$\simeq 7.90$, cf.\ Fig.~\ref{fig2}, right panel). 
Other mechanisms or sources are needed to explain the origin of nebular \heii\ \citep[see, e.g.,][and references therein]{olivier2021, simmonds2021}.

At $z \sim 2$, \cite{Naidu2021} recently identified two groups of LAEs, one probably showing strong LyC escape, and the
other low LyC escape fractions. Comparing the stacked spectra of these two groups, they found differences in their observed rest-frame UV spectra, with the strong leaker candidates showing high-ionization lines of C~{\sc iv}~1550\AA\ and He~{\sc ii}~1640\AA, and the other group showing only lower-ionization lines (e.g.,\ C~{\sc iii} 1909\AA).
From this finding, \cite{Naidu2021} suggested that strong LyC leakers could have harder ionizing spectra. 
On the other hand, if the galaxies studied in our work are comparable to high-$z$ LAEs, our results imply that these observed differences are not related to LyC escape.

Generally speaking, the absence of hardness variations with \fesc\ shows that this property of the global radiation field does not determine the conditions for the ionizing photon escape. In contrast, other studies have found significant correlations between different physical properties and \fesc, which might indicate such physical processes. This includes highly concentrated star formation (high SFR surface densities), a high ionization parameter, an inhomogeneous ISM and dust distribution, and low amounts of dust \citep{Verhamme2017Lyman-alpha-spe,gazagnes2018,Cen2020Physics-of-Prod,Flury2022b,Saldana2022}. 
Our finding does not exclude that radiative processes contribute to determining LyC escape, but it indicates that the hardness of the radiation field (over the energy range measured here) is not fundamental.
For example, this suggests that LyC escape is not related to low-luminosity active galactic nuclei (AGNs).
However, our differential study does not explain the origin of nebular \heii\ emission, which is known to require sources of ionizing photons above 54 eV in amounts that are not predicted for normal stellar populations \citep[see, e.g.,][]{shirazi2012,Stasinska2015Excitation-prop,schaerer2019}.

\section{Summary}

We have empirically investigated the hardness of the ionizing spectra between 54.4 and 13.6 eV (\qrat, i.e.,\ the ionization potential of He$^+$ and that of H) as a function of the LyC escape fraction, $f_{\rm esc} \rm (LyC)$, of a large sample of star-forming galaxies at low redshift for which LyC are available from {\em HST} observations. 
Optical recombination lines of \heii~4686\AA{ }and \hb~4861\AA{ }from SDSS and XShooter spectra were used to determine \qrat\ and its dependence on $f_{\rm esc} \rm (LyC)$. The underlying effect of metallicity was also considered. 

We built stacked spectra in bins of $f_{\rm esc} \rm (LyC)$ and metallicity, allowing us to study the behavior of \heii/\hb\ across a wide range of $f_{\rm esc} \rm (LyC) \simeq 0-0.9$ and \oh~$=7.5-8.6$. 
We find that the intensity of \heii/\hb\ does not depend on $f_{\rm esc} \rm (LyC)$. Very strong leakers (<$f_{\rm esc} \rm (LyC)$>~$\simeq 0.5$) and nonleakers (<$f_{\rm esc} \rm (LyC)$>~$\simeq 0$) have similar intensities of \heii/\hb\ on average, about \heii/\hb~$\simeq 0.01-0.02$. Instead, we find that \heii/\hb\ is primarily driven by variations in metallicity, where \heii/\hb\ increases with decreasing \oh, in particular, for \oh~$<8.0$, as known from previous studies \citep[e.g.,][]{shirazi2012,schaerer2019}. 
At comparable metallicities, strong LyC leakers present roughly the same \heii/\hb\ intensities as non- or weak leakers and as in many other normal star-forming galaxies, where nebular \heii\ is detected.

In short, our results demonstrate that \qrat\ does not correlate with $f_{\rm esc} \rm (LyC)$. This implies that strong LyC emitters do not show harder ionizing spectra than nonleakers at similar metallicity. 
Future studies will address other hardness or softness indicators of the radiation field and will more broadly examine the nebular properties of the galaxies from the LzLCs.

\begin{acknowledgements}
The authors thank the referee for useful comments that greatly improved the clarity of this work. 
Y.I.I. acknowledges support from the National Academy of Sciences of Ukraine by its priority project “Fundamental properties of the matter in the microworld, astrophysics and cosmology”. R.A. acknowledges support from ANID Fondecyt Regular Grant 1202007. 
Funding for the Sloan Digital Sky Survey IV has been provided by the 
Alfred P. Sloan Foundation, the U.S. Department of Energy Office of 
Science, and the Participating Institutions. 
SDSS-IV acknowledges support and resources from the Center for High 
Performance Computing  at the University of Utah. The SDSS website is www.sdss.org.

\end{acknowledgements}


\bibliographystyle{aa}
\bibliography{main}

\end{document}